\documentclass{ptapap}

\author{B.\,Buysschaert}[bram.buysschaert@obspm.fr, LESIA,IvS]
\author{C.\,Neiner}[LESIA]
\author{T.\,Ramiaramanantsoa}[MONT]
\author{N.D.\,Richardson}[TOL]
\author{A.\,David-Uraz}[FLOR]
\author{A.F.J.\,Moffat}[MONT]
\author{BEST}[]
\affil[LESIA]{LESIA, Observatoire de Paris, PSL Research University, CNRS, Sorbonne Universit\'es, UPMC Univ. Paris 06, Univ. Paris Diderot, Sorbonne Paris Cit\'e, 5 place Jules Janssen, F-92195 Meudon, France}
\affil[IvS]{Instituut voor Sterrenkunde, KU Leuven, Celestijnenlaan 200D, 3001 Leuven, Belgium}
\affil[MONT]{D\'epartement de physique and Centre de Recherche en Astrophysique du Qu\'ebec (CRAQ), Universit\'e de Montr\'eal, C.P. 6128,
Succ. Centre-Ville, Montr\'eal, Qu\'ebec, H3C 3J7, Canada}
\affil[TOL]{Ritter Observatory, Department of Physics and Astronomy, The University of Toledo, Toledo, OH 43606-3390, USA}
\affil[FLOR]{Department of Physics \& Space Sciences, Florida Institute of Technology, Melbourne, FL, 32901, USA \\ * Based on data collected by the BRITE Constellation satellite mission, designed, built, launched, operated and supported by the Austrian Research Promotion Agency (FFG), the University of Vienna, the Technical University of Graz, the Canadian Space Agency (CSA), the University of Toronto Institute for Aerospace Studies (UTIAS), the Foundation for Polish Science \& Technology (FNiTP MNiSW), and National Science Centre (NCN).\\ ** Based on CHIRON spectra collected under CNTAC proposal CN2015A-122.}

\title{Understanding the photometric variability of $\zeta$\,Ori\,Aa* **}

\begin{document}

\maketitle

\begin{abstract}
We studied the variability of the magnetic O-type supergiant$\zeta$\,Ori\,Aa using multi-colour BRITE photometry.  We confirmed the known rotation frequency $f_{\rm rot} = 0.15 \pm 0.02$\,c/d, and detected some of its higher harmonics, of which $4f_{\rm rot}$ is compatible with the known DAC recurrence timescale.  Thanks to simultaneous high-resolution CHIRON spectroscopy, we could identify another frequency $f_{\rm env} = 0.10 \pm 0.02$\,c/d, caused by the circumstellar environment.  Variations in the circumstellar environment are believed to cause the observed difference between the BRITE lightcurves.

\end{abstract}

Massive stars show various types of time-dependent variability.  This variability can originate at the stellar surface, e.g. non-radial stellar pulsations or surface brightness inhomogeneities, or it can be related to the circumstellar environment, e.g. Discrete Absorption Components (DACs) - related to Corotating Interaction Regions (CIRs) - or magnetospheres .  In this work, we studied $\zeta$\,Ori\,Aa, which is the only known magnetic O-type supergiant as of today.

\section{$\zeta$\,Orionis}
$\zeta$\,Ori is a hierarchical multiple system, comprising of $\zeta$\,Ori\,A and $\zeta$\,Ori\,B in a visual binary system.  The \textit{Washington Double Star Catalog} indicated an orbital period of about 1509\,y with an eccentricity of 0.07 using the common proper motion \citep{2001AJ....122.3466M}.  More recently, \citet{2013A+A...554A..52H} characterized the orbit of the Aa+Ab subsystem.  This system is slightly eccentric ($e=0.338\pm0.004$) with an orbital period of approximately 7.4\,y.  $\zeta$\,Ori\,Aa has an O9.2Ib\,N-weak spectral type, while $\zeta$\,Ori\,B and $\zeta$\,Ori\,Ab are classified as B0III and B1V, respectively \citep{2013A+A...554A..52H, 2014ApJS..211...10S}.

A weak magnetic field was detected for the massive supergiant $\zeta$\,Ori\,A \citep{2008MNRAS.389...75B} using high-resolution spectropolarimetry. \citet{2015A+A...582A.110B} refined the study of the magnetic field of the supergiant by accounting for $\zeta$\,Ori\,Ab and additional observations.  They determined the rotation period of $\zeta$\,Ori\,Aa ($P_{\rm rot} = 6.83 \pm 0.08$\,d) and confirmed the presence of a weak dipolar magnetic field ($B_d \approx 140$\,G).  Using spectral disentangling, $\zeta$\,Ori\,Ab was isolated in the observations, leading to the non-detection of a magnetic field for the secondary, with a detection threshold of 300\,G.  The authors also found additional variability in their 2011--12 magnetic measurements compared to those of the 2007--8 campaign.

Finally, \citet{1996A+AS..116..257K, 1999A+A...344..231K} detected weak DACs for $\zeta$\,Ori\,A using UV spectroscopy.  The analysis was hampered by a short timebase and a poor cadence, yet they were able to determine the recurrence timescale of the DACs to be $t_{\rm rec} = 1.6\pm 0.2$\,d, about 1/4 of the rotation period.

\section{BRITE photometric study}
\subsection{Observations}
The BRIght Target Explorer (BRITE) \citep{2014PASP..126..573W, 2016arXiv160800282P} monitored $\zeta$\,Ori during two separate observing campaigns, each time using both red and blue nano-satellites.  All observations were taken in 'classical' mode with a typical exposure of 1\,s.  However, on-board stacking was applied for some of the measurements.  Lightcurves were extracted from the downlinked pixel rasters using circular apertures (Popowicz et al., in prep.) and additional meta-data (CCD centroid positions and CCD temperature) were provided.  We developed specific routines to further correct the extracted BRITE photometry (Buysschaert et al. 2016, these proceedings; Buysschaert et al. in prep.).  We corrected the timing of the observations, cleaned the data by performing outlier rejection, reduced the noise by accounting for the variable PSF shape, and corrected for instrumental effects.  Lastly, we calculated the duty cycles and root-mean-square (RMS) of the flux to assess the quality of the BRITE photometry.  Only the best data was considered for further analysis.  We show the data characteristics in Table\,\ref{tab:photometry} and present the individual lightcurves in Fig.\,\ref{fig:lightcurves}.  Simultaneous red and blue photometry agree well.  Since $\zeta$\,Ori\,Aa is the brightest component in the system, we assume in the following analysis that all BRITE variability is caused by $\zeta$\,Ori\,Aa or its circumstellar environment.

\begin{table}[t]
\caption{Diagnostics related to studied BRITE observations of $\zeta$\,Ori.  We mark the satellite name, the visit and setup number, the number of on-board stacked observations, the BRITE epoch  as defined in this work, the start and length of the observations, the RMS of the corrected flux, and the satellite duty cycle.}
\centering
\tabcolsep=6pt
\begin{tabular}{p{1cm}lllp{1.3cm}lllll}
\hline
\hline
Satellite & Run & Setup & Stacked & Epoch & $\mathrm{T_{start}}$ & Length & $\mathrm{RMS_{corr}}$ & $\mathrm{D_{sat}}$\\
&&Set&Observ.&& [HJD& [d] & [ppt] & [\%]\\
&&&&&-2450000] & &  & \\
\hline
BAb			&	I	&	3	&	1	&	Orion\,I		&	6628.43	&	74.01	&	0.86	&	32.2\\
BAb			&	I	&	4	&	1	&	Orion\,I		&	6702.51	&	29.99	&	0.72	&	66.5\\
UBr			&	I	&	7	&	1	&	Orion\,I		&	6603.61	&	129.28	&	0.64	&	40.6\\\\
															
BLb			&	II	&	3	&	5	&	Orion\,IIa	&	6998.52	&	45.04	&	1.19	&	75.8\\
BLb			&	II	&	6	&	1	&	Orion\,IIb  &	7052.75	&	45.53	&	0.55&	89.2\\
BHr			&	II	&	5	&	5	&	Orion\,IIa  &	6998.56	&	49.06	&	0.85	&	78.1\\
BHr			&	II	&	7	&	1	&	Orion\,IIb  &	7056.99	&	38.53	&	0.52	&	67.7\\
\hline
\end{tabular}
\label{tab:photometry}
\end{table}

\subsection{Time-series analysis}
Since the frequency diagrams of $\zeta$\,Ori's BRITE lightcurves indicate that the character of their variability differs, we subdivided this data into three distinct epochs, namely Orion\,I, Orion\,IIa, Orion\,IIb.

\begin{figure}
\includegraphics[width=\textwidth]{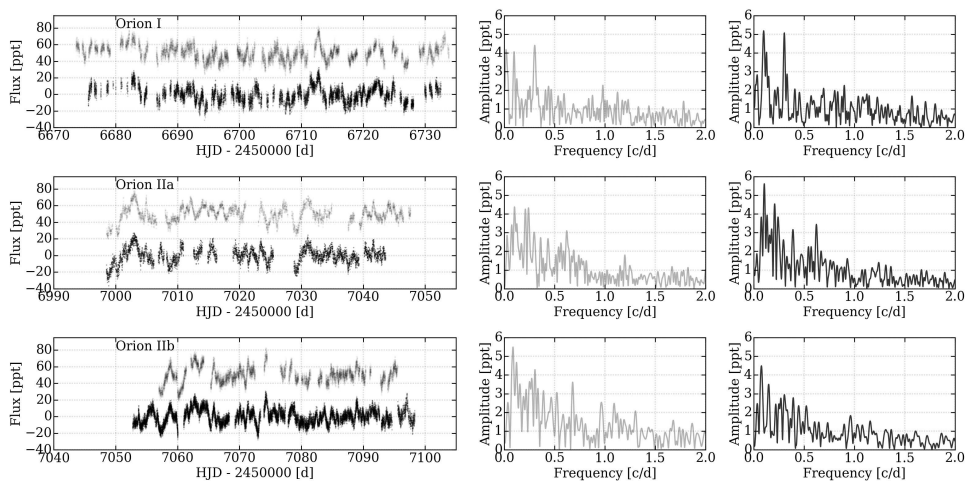}
\caption{Studied BRITE lightcurves for $\zeta$\,Ori separated per epoch (left panels) with their corresponding Lomb-Scargle periodograms (middle and right panels).  The blue flux photometry is given in black, while the red lightcurves are marked in gray and have an offset of 50 parts-per-thousand (ppt) for enhanced visibility. No significant variability was found above 1\,c/d.}
\label{fig:lightcurves}
\end{figure}

As a first step, we performed an iterative prewhitening approach to determine the significant periodic variability present in each lightcurve.  We used ten times oversampled Lomb-Scargle periodograms in the frequency domain up to 10\,c/d.  These frequency diagrams are presented in Fig.\,\ref{fig:lightcurves}.  During the detailed frequency analysis, we considered a frequency peak significant if it reached a SNR greater than four within a frequency window of 4\,c/d.  Each of the extracted frequencies was then compared between the various studied BRITE lightcurves and only considered for further inspection if it was obtained in at least two lightcurves.

\begin{figure}
\includegraphics[width=\textwidth]{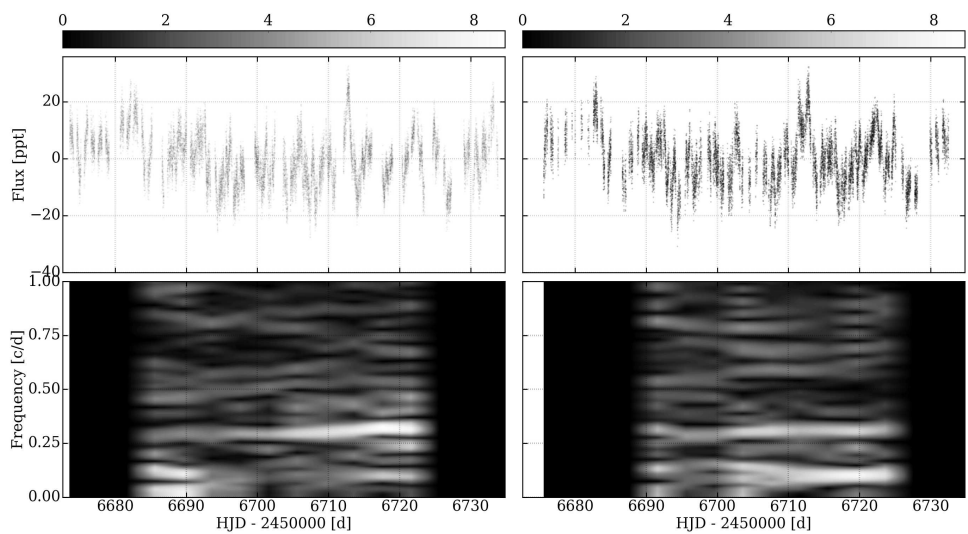}
\caption{STFTs of the Orion\,I data (bottom panels), compared to their respective lightcurves (top panels).  The amplitude of the variations within the STFT are indicated by the gray scale.  Red photometry is shown on the left, and blue photometry on the right.}
\label{fig:STFT}
\end{figure}

Second, we determined the short-time Fourier transform (STFT) to investigate the variability of the periodic fluctuations with respect to time.  Since most of the variability happens in the low frequency regime, a sufficiently high frequency resolution was needed.  Therefore, a time window of 20\,d (leading to $\delta f = 0.05$\,c/d) was chosen, together with a timestep of 4\,d.  We again employed ten times oversampled Lomb-Scargle periodograms.  No periodograms were calculated when the duty cycle was below 50\,\%.  An example of such SFTF is shown in Fig.\,\ref{fig:STFT}.

\subsection{Results}
Performing an iterative prewhitening approach on all BRITE lightcurves of $\zeta$\,Ori and comparing the extracted frequencies lead to 17 significant frequencies.  These are understood to be simple linear combinations and higher harmonics of 3 individual frequencies, namely $f_{\rm rot} = 0.15 \pm 0.02$\,c/d; $f_{\rm env} = 0.10 \pm 0.02$\,c/d; and $f_{\rm x} = 0.03 \pm 0.02$\,c/d.  We discuss these further in Sect.\,\ref{sec:discussion}.  The STFT analysis confirmed the differences between the various BRITE epochs for $\zeta$\,Ori.  In addition, the amplitudes of variability changes with time within a given dataset and between the various epochs, and sharp features are observed, indicating that non-periodic events could also be present in the observations.

\section{CHIRON spectroscopic study}
\subsection{Observations}
$\zeta$\,Ori was observed 65 times, typically twice per night, between 06/02/2015 and 16/03/2015, simultaneously with BRITE, by the CHIRON \'{e}chelle spectrograph mounted on the 1.5m-CTIO telescope \citep{2013PASP..125.1336T}.  The spectroscopic observations include the 459\,nm -- 760\,nm range, with a wavelength resolution of 80000, and is the combination of two consecutive 8-s sub-exposures.  These observations were corrected with the CHIRON pipeline for bias and flat-field effects and wavelength calibrated.  Lastly, they were blaze-corrected and normalized to the continuum level.

\subsection{Time-series analysis}
\begin{figure}
\begin{centering}
\includegraphics[width=0.5\textwidth]{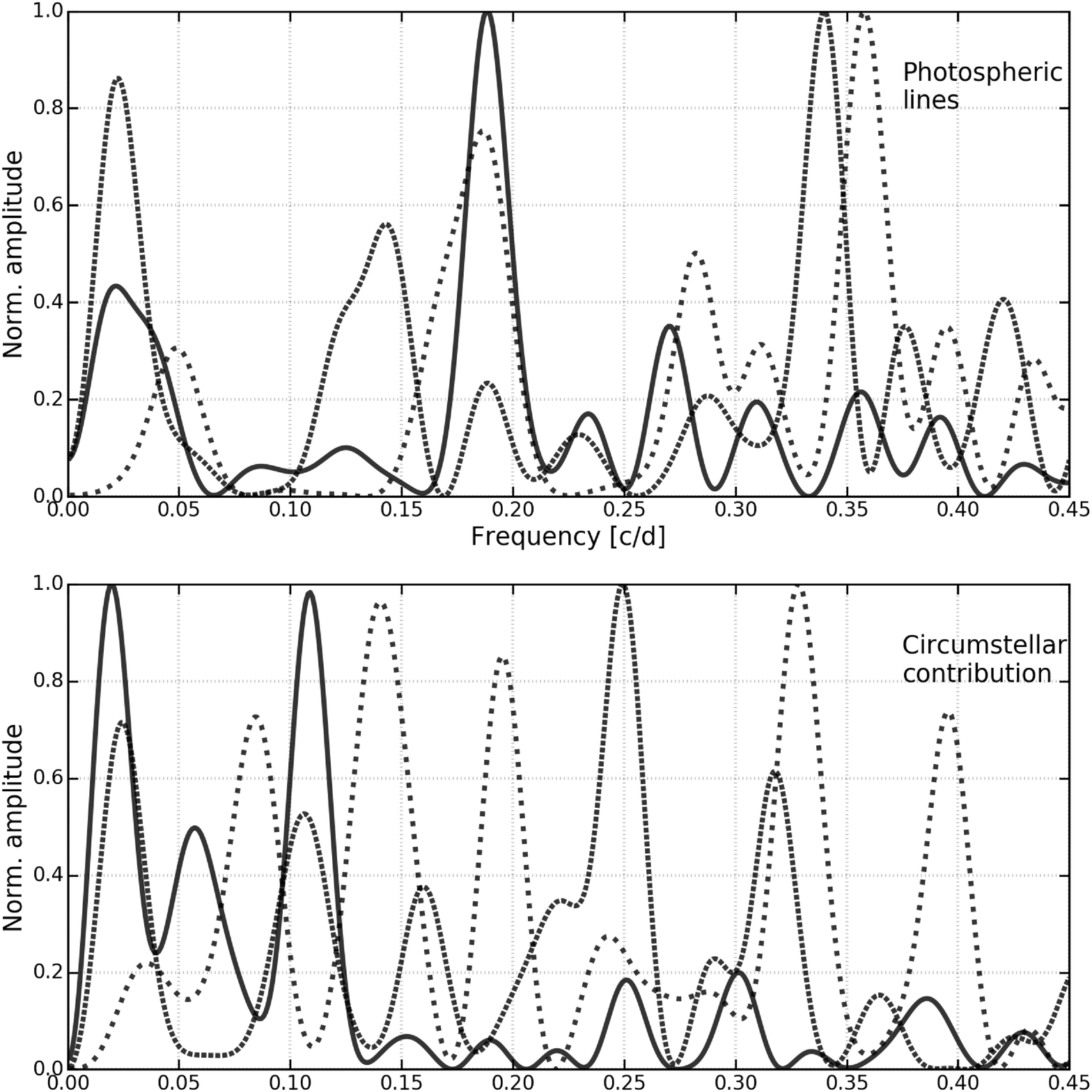}
\caption{Normalized CLEAN periodograms showing the variations of selected spectroscopic lines. \textit{Top}: solid - He\,\textsc{II}\,$\lambda\lambda5411.5$; dashed - O\,\textsc{III}\,$\lambda\lambda5592.3$; dotted - C\,\textsc{IV}\,$\lambda\lambda5801.3$.  \textit{Bottom}: solid - H$\alpha$; dashed - He\,\textsc{I}\,$\lambda\lambda6678.2$; dotted - He\,\textsc{I}\,$\lambda\lambda4921.9$.}
\label{fig:EW}
\end{centering}
\end{figure}

We selected several spectroscopic lines known either to be influenced by the circumstellar environment (H$\alpha$, He\,\textsc{I}\,$\lambda\lambda6678.2$, and He\,\textsc{I}\,$\lambda\lambda4921.9$) or either to be purely photospheric (He\,\textsc{II}\,$\lambda\lambda5411.5$, O\,\textsc{III}\,$\lambda\lambda5592.3$, and C\,\textsc{IV}\,$\lambda\lambda5801.3$).  To study the time-dependent variability of these lines, we measured their respective equivalent width (EW).  For H$\alpha$, we performed this EW integration over both the emission and absorption feature of the P\,Cygni profile.  Next, we calculated CLEAN periodograms of the EW variations \citep{1987AJ.....93..968R}, using a gain of 0.5 and 10 iterations, because this type of periodograms is less influenced by the moderate sampling.  We limited the investigated frequency domain to 0.5\,c/d, again to be less influenced by aliasing effects.  We show these periodograms in Fig.\,\ref{fig:EW}.

\subsection{Results}
The periodograms of the EW variations for lines with a circumstellar contribution and those purely photospheric look fairly different.  However, the data did not permit a detailed frequency analysis.  Therefore, we compared these periodograms to those of the simultaneous BRITE photometry (Orion\,IIb).  For circumstellar environment polluted lines, the dominant variability occurs in the region of $f_{\rm env}=0.10\pm0.03$\,c/d, with the He\,I lines showing some additional variability with $f_{\rm rot}=0.15\pm0.003$\,c/d and $2f_{\rm rot}$.  The purely photospheric lines, however, exhibit their strongest variability with a frequency of $0.19\pm0.03$\,c/d, as well as with $2f_{\rm rot}$.

\section{Discussion}
\label{sec:discussion}
\subsection{Rotation}
For each studied BRITE lightcurve, we determined variability related to the rotation frequency and its higher harmonics.  Our photometric study of the BRITE data indicates $f_{\rm rot} = 0.15 \pm 0.02$\,c/d, compatible with the literature value for $\zeta$\,Ori\,Aa of $0.146 \pm 0.002$\,c/d \citep{2015A+A...582A.110B}.  The strongest variability occurs with $2f_{\rm rot}$, as often observed for hot magnetic stars because of their dipole field.

\subsection{DACs}
Another harmonic of the rotation frequency, $4f_{\rm rot}$, was often retrieved.  This variability coincides with the literature DAC recurrence timescale with $f_{\rm rec} = 0.625\pm0.075$\,c/d \citep{1999A+A...344..231K}.  DACs are understood to be related to surface inhomogeneities connected to CIRs.  Yet the exact mechanism to produce and sustain four (enhanced) brightness regions on $\zeta$\,Ori\,Aa remains uncertain.  Both the STFT and the individual lightcurves indicate that the amplitude varies with time.

\subsection{Circumstellar environment}
It is thanks to the CHIRON spectroscopy that we were able to comprehend the photometric variability of $f_{\rm env} = 0.10 \pm 0.02$\,c/d, as this variation is only present in spectroscopic lines with a circumstellar contribution, while being strikingly absent in purely photospheric lines.  Therefore, we conclude that this variability originates in the circumstellar environment of  $\zeta$\,Ori\,Aa.  This variability is to not related to the stellar rotation, hence it is not produced by the weak magnetosphere of $\zeta$\,Ori\,Aa.  Instead, we propose periodic mass loss as the driving mechanism for this variability, possibly produced by beating between undetected non-radial pulsations, similar to the Be-phenomenon \citep[e.g.][]{1998cvsw.conf..207R, 2003A+A...411..229R}.  As the circumstellar environment influences the BRITE photometric measurements, it could naturally explain the differences observed between the studied BRITE epochs.

\subsection{Low-frequency variability}
To explain all the observed frequency combinations, we need one additional independent frequency (besides $f_{\rm rot}$ and $f_{\rm env}$).  We chose $f_{\rm x} = 0.03 \pm 0.02$\,c/d, as it produces the combinations in the easiest way.  Yet, no easy physical interpretation is available for this variability.  We note that it is close to the frequency precision.  In addition, the power of this frequency was significantly altered during the correction of the extracted BRITE lightcurves.  Therefore, we conclude that this frequency might not be physical.

\section{Conclusions}
For the magnetic massive supergiant $\zeta$\,Ori\,Aa, we performed a detailed time-series analysis of the individual red and blue BRITE photometry, as well as of the simultaneous high-resolution CHIRON spectroscopy.  This study identified three independent frequencies, two of stellar origin and one likely non-physical, leading to several higher harmonics and simple linear frequency combinations.

Our value for the rotation frequency is compatible with the literature value.  We have also detected variability with $2f_{\rm rot}$, corresponding to the magnetic dipole, and $4f_{\rm rot}$, agreeing with the known DAC recurrence timescale.  We propose that the source of the DACs is located at the stellar surface and lay at the origin of the CIRs.  With simultaneous CHIRON spectroscopy, we determined the circumstellar environment to be variable with a frequency of $0.10 \pm 0.02$\,c/d.

\bibliographystyle{ptapap}
\bibliography{PhD_ADS}

\end{document}